\documentclass[pmlr]{jmlr}

\RequirePackage{graphicx}
 \usepackage{booktabs}
\usepackage{longtable}
\usepackage{algpseudocode}

\makeatletter
\def\set@curr@file#1{\def\@curr@file{#1}} 
\makeatother
\usepackage[load-configurations=version-1]{siunitx} 

\newcommand{\cs}[1]{\texttt{\char`\\#1}}

\theorembodyfont{\upshape}
\theoremheaderfont{\scshape}
\theorempostheader{:}
\theoremsep{\newline}

\jmlrvolume{[VOLUME \# TBD]}
\jmlryear{2024}
\jmlrworkshop{Machine Learning for Healthcare}

\title[Profiling Copy Number Variations]{Classifying Copy Number Variations Using State Space Modeling of Targeted Sequencing Data: A Case Study in Thalassemia}

\author{\Name{Austin Talbot}
       \Email{talbota@pillarbiosci.com}\\ 
       \addr Pillar Biosciences\\
       Natick, MA, USA 
       \AND
       \Name{Alex V. Kotlar}
       \Email{akotlar@bystro.io}\\ 
       \addr Bystro AI\\
       Boston, MA, USA 
       \AND
       \Name{Lavanya Rishishwar}
       \Email{rishishwarl@pillarbiosci.com}\\ 
       \addr Pillar Biosciences\\
       Natick, MA, USA 
       \AND
       \Name{Yue Ke}
       \Email{key@pillarbiosci.com}\\ 
       \addr Pillar Biosciences\\
       Natick, MA, USA 
       }

\begin{document}

\maketitle

\begin{abstract}
Thalassemia, a blood disorder and one of the most prevalent hereditary genetic disorders worldwide, is often caused by copy number variations (CNVs) in the hemoglobin genes. This disorder has incredible diversity, with a large number of distinct profiles corresponding to alterations of different regions in the genes. Correctly classifying an individual's profile is critical as it impacts treatment, prognosis, and genetic counseling. However, genetic classification is challenging due to the large number of profiles worldwide, and often requires a large number of sequential tests. Targeted next generation sequencing (NGS), which characterizes segments of an individual's genome, has the potential to dramatically reduce the cost of testing and increase accuracy. In this work, we introduce a probabilistic state space model for profiling thalassemia from targeted NGS data, which naturally characterize the spatial ordering of the genes along the chromosome. We then use decision theory to choose the best profile among the different options. Due to our use of Bayesian methodology, we are also able to detect low-quality samples to be excluded from consideration, an important component of clinical screening. We evaluate our model on a dataset of 57 individuals, including both controls and cases with a variety of thalassemia profiles. Our model has a sensitivity of $0.99$ and specificity of $0.93$ for thalassemia detection, and accuracy of $91.5\%$ for characterizing subtypes. Furthermore, the specificity and accuracy rise to $0.96$ and $93.9\%$ when low-quality samples are excluded using our automated quality control method. This approach outperforms alternative methods, particularly in specificity, and is broadly applicable to other disorders.

\end{abstract}

\section{Introduction}\label{sec:introduction}

Thalassemia is a genetic blood disorder that disrupts the production of hemoglobin, resulting in chronic anemia and other health complications \cite{weatherall2001inherited}. It is the most prevalent genetic disorder worldwide and poses a significant public health challenge, particularly in regions such as the Mediterranean, Southeast Asia, and sub-Saharan Africa \cite{weatherall2008thalassaemia,modell2008global}. Thalassemia is caused by copy number variations (CNVs) or single nucleotide variants (SNVs) in the \textit{HBA} (alpha-globin) and \textit{HBB} (beta-globin) genes. In CNV-caused thalassemia, while individuals normally possess four copies of each gene, affected individuals have alterations in portions of one or more copies. The severity of anemia depends on the number of missing copies, ranging from mild symptoms to life-threatening conditions \cite{galanello2011alpha,cao2010beta,rund2005beta}. The portions of the genes that are altered can differ dramatically between individuals due to the diverse geographic origins of the disease. Accurate characterization of these ``profiles'' (specific patterns of alterations in the gene) is important for treatment and genetic counseling.

Traditional diagnostic methods rely on a sequence of hematologic assessments combined with genetic testing. Preliminary screening involves a complete blood count, which often reveals anemia with low mean corpuscular volume and an elevated red blood count. A peripheral blood smear may further support the diagnosis by showing microcytosis, hypochromia, target cells, and basophilic stippling. Hemoglobin electrophoresis or high-performance liquid chromatography (HPLC) is then used to quantify hemoglobin fractions. In $\beta$-thalassemia, elevated hemoglobin A2 (HbA2) and sometimes hemoglobin F (HbF) are observed, whereas $\alpha$-thalassemia often results in normal HbA2 and HbF levels, making it more difficult to detect without molecular testing \cite{ghosh2014guidelines,adekile2015clinical}. Iron studies are typically conducted to rule out iron deficiency anemia \citep{galanello2011alpha}. 

Next-generation sequencing (NGS) has transformed diagnosis of genetic disorders,  allowing a small number of tests to completely characterize the genetic profile of the individual. In NGS, an individual’s DNA is fragmented, amplified, and sequenced in parallel, allowing for the reconstruction of genomic sequences. While whole-genome sequencing (WGS) reconstructs the entire genome for a more complete characterization of disease status, targeted sequencing—where only specific regions of interest are amplified—provides a cost-effective alternative for disease diagnosis \cite{rehm2013disease,ku2013next}. By designing amplicons that capture clinically relevant regions, targeted sequencing minimizes unnecessary data collection and optimizes throughput in clinical settings \cite{mamanova2010target,rabbani2012next}.


Unfortunately, detecting CNVs, such as thalassemia, from NGS data remains a significant challenge \cite{alkan2011genome}. Unlike single-nucleotide variants (SNVs), which can be identified by sequence mismatches, CNV detection relies on quantifying read-depth differences relative to a reference sample (clinical normal). This process is highly susceptible to variability introduced by sequencing biases, amplification artifacts, and technical noise \cite{teo2012statistical}. These difficulties are amplified in thalassemia profiling, which requires classification of the correct genetic profile of the thalassemia mutation. Existing CNV detection methods often require either (i) a large number clinical normal samples \cite{krumm2012copy,backenroth2014canoes},  which is impractical in clinical settings, or (ii) CNVs spanning large genomic regions \cite{talevich2016cnvkit,d2016enhanced}, which is not well suited for thalassemia where deletions can be small and localized. Moreover, current publicly-available approaches focus solely on identifying the existence of CNVs, the subsequent classification task is not included. 

In this paper, we introduce a novel probabilistic approach for jointly estimating CNVs and  classification of mutation types. While this is applicable to profiling a wide class of genetic disorders using targeted sequencing, we demonstrate its performance on thalassemia specifically. Our method is based on state-space modeling \cite{chopin2020introduction}, a framework well-suited for sequential data, as amplicons are spatially ordered along the genome. This allows for more accurate copy number estimation by leveraging correlations between adjacent amplicons, while allowing for sudden transitions corresponding to deletions or amplifications. Using these estimates, we then apply Bayesian decision theory to assign each sample to a predefined thalassemia profile, accounting for both genetic variability and prior knowledge. Finally, we are able to evaluate sample quality via the Bayesian evidence of the state space model, quantifying the relative confidence in the model fit of the data. Our framework offers several advantages, namely (i) an estimation of copy number at each targeted region of the gene, refining CNV calls beyond simple presence/absence detection, (ii) a probabilistic classification of thalassemia profile, incorporating genetic inheritance patterns, and (iii) automatic sample quality evaluation using Bayesian evidence, allowing clinicians to assess confidence levels in each prediction, enhancing clinical interpretability. To properly implement these methods, we require altering standard inference of the state space model using sequential Monte Carlo in order to handle the abrupt changes in copy number. We evaluate the performance of our model on a mixture of synthetic and real data, and show that our method has high sensitivity and specificity, especially when (iii) is used to reject low-quality samples.


\subsection*{Generalizable Insights about Machine Learning in the Context of Healthcare}

\begin{itemize}
    \item Bayesian evidence significantly enhances diagnostic accuracy by identifying low-quality samples that require manual review or rerunning.
    \item Auxiliary particle filtering improves performance in state space models, particularly in scenarios where abrupt changes are expected.
\end{itemize}

\section{Related Work}\label{sec:relatedwork}
There are three areas of related work; methods developed to detect thalassemia using NGS, methods for CNV quantification, and work with state space modeling of genomic data.

Traditional thalassemia profiling requires multiple assays: Sanger sequencing for SNVs and small indels, and GAP-PCR, MLPA, or chromosomal microarrays for structural variants (SVs). NGS offers a comprehensive alternative, but early amplicon-based approaches targeted only specific SVs, requiring pooled PCR reactions. A more costly hybrid-capture NGS assay covering 275 kb of hemoglobin-related regions was introduced by \citet{shang2017rapid}. More recently, \citet{cao2022ngs4thal} developed NGS4THAL, a bioinformatics tool for thalassemia variant detection, but it depends on whole-genome or hybrid-capture sequencing for continuous coverage. Despite these advances, clinical integration of NGS remains limited due to complex library preparation and the computational demands of SNV and SV detection.

The literature on more generic CNV estimation is quite extensive \cite{gordeeva2021benchmarking}. Many of these methods, such as CNVkit \cite{talevich2016cnvkit} and Excavator \cite{magi2013excavator}, are optimized for detecting large-scale copy number variations (CNVs) and typically require large sample cohorts for normalization. CNVkit is designed for targeted and exome sequencing but relies on pooled normal samples or panel-based reference creation, making it less effective in small-cohort or single-sample settings. Excavator, developed for whole-exome and whole-genome sequencing, employs a hidden Markov model (HMM) to detect CNVs but assumes broad genomic alterations rather than small, localized deletions, such as those characteristic of thalassemia. Other tools, including GATK-gCNV \cite{babadi2023gatk} and CoNIFER \cite{krumm2012copy}, similarly depend on large reference panels to reduce noise and improve sensitivity, making them impractical for clinical applications where individual patient testing is required. Moreover, while these methods estimate CNVs, none provide a subsequent classification step that assigns the detected copy number changes to predefined disease-specific profiles. Due to the shortcomings of the previously described methods, to our knowledge our algorithm has no publicly-available competitors.

The use of state space models applied to genomics is quite extensive. Early work with hidden Markov models (HMMs) demonstrated their utility for large CNV detection \citep{scharpf2008hidden,xu2011genome}. More general state-space approaches for genomic segmentation have been proposed \cite{zhu2012signal}. These models have substantial advantages over traditional CNV segmentation methods due to their explicit modeling of uncertainty and flexibility to adapt to customizable observational models. These models often depend on Markov chain Monte Carlo methods (specifically particle filtering) \citep{chopin2020introduction}. Among this class of algorithms, auxiliary particle filtering \citet{douc2011sequential} is the most pertinent to this work, as we use this technique to address difficulties associated with the abrupt changes occuring in the series. 


\section{Methods}\label{sec:methods}
\subsection {Notation and Data}

The data come from targeted amplicon sequncing. In this type of NGS, small sections of the DNA, known as amplicons, are repeatedly copied (amplified)  using a process called PCR. After amplification, a subset of these copies are read using a sequencing machine and the number of times a particular amplicon is read is the read count. More copies of a gene in the original starting material will result in higher read counts, with noise due to both  amplification and sequencing.

We analyze read counts from $J$ targeted amplicons, of which $K=\{i_1,\dots,i_K\}$ are control amplicons coming from regions known to be CNV neutral. The counts are obtained from the case sample, denoted as $\{s_j\}_{j=1:J}$, and a corresponding normal sample, $\{n_j\}_{j=1:J}$. To assess copy number variations (CNVs), we compute the log copy ratio for each amplicon $\tilde{y}_j=\log(s_j+10)-\log(n_j+10)$ (the offset of $10$ is used to handle the case of homozygous deletions). Functionally, the purpose of the normal sample is to account for amplicon-specific effects on the read counts. Often, there are systematic biases in read counts between the sample and normal stemming from different sequencing depths. We use the control amplicons to account for this by subtracting the mean of the log ratios in the control amplicons, as 
\begin{equation}
    y_j=\tilde{y}_j - Mean(\tilde{y}_{K}).
\end{equation}
After this transformation, $y_j = 0$ indicates a copy-neutral amplicon (identical to the control amplicons), while negative values indicate deletions and positive values indicate amplifications. We refer to these covariates as the log copy number ratio (lCNR), as these quantities ideally are the log of the ratio of copy numbers between the sample and the control.

Thalassemia profiling relies on clinically predefined  ``profiles'', each representing expected log ratios in regions in the genome (and by extension the amplicons). We denote the expected log ratio for the $i$-th profile as $d_i = \{d_{ij}\}_{j=1:J}$. These profiles vary based on population-specific mutations (i.e. Southeast Asian  or Turkish deletions). These profiles may correspond to heterozygous or homozygous deletions and amplifications. For a given sample we must estimate the number of copies at each amplicon, and subsequently identify the most likely thalassemia profile for both the alpha and beta genes. Example profiles are shown in the top right of Figure \ref{figmethod}.

\subsection{A State Space Model of the Log Copy Ratios}

\begin{figure}[ht]
\centering
\includegraphics[width=1.0\textwidth]{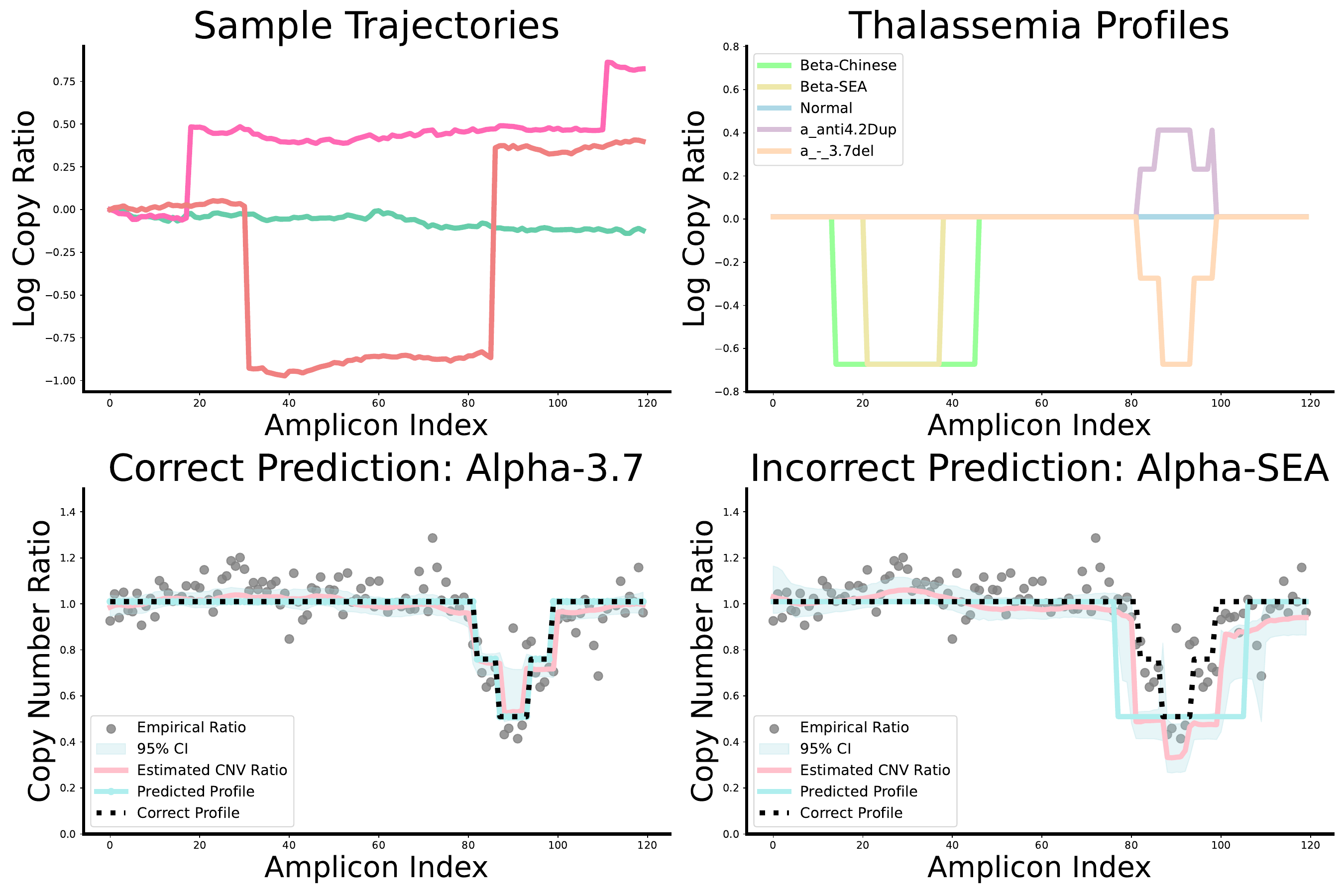}
\caption{The top left shows sample trajectories drawn from \ref{eq:transition_kernel} with $\sigma^2=1$, $p=0.02$, and $\epsilon=.001$. The top right shows the profiles of 5 different thalassemia profiles, including two beta deletions, an alpha deletion, and an alpha duplication. The bottom left shows a sample predicted correctly by our model, including the posterior mean, credible interval, and diagnosis. The bottom right is identical, except the model predicted south-east Asian deletion (SEA) rather than the correct profile alpha 3.7 deletion.}\label{figmethod}
\end{figure}

The lCNRs exhibit strong spatial correlation due to the physical ordering of the amplicons along the chromosome. A natural model for this dependency is through a state-space model \cite{chopin2020introduction}. According to these models,  the observed lCNRs are noisy measurements of the true lCNR. Correlations in these unobserved values are induced via a Markov chain of the true lCNRs. Formally, denoting the unobserved true lCNR as $x_j$, we model this chain as 
\begin{equation}\label{eq:transition_kernel} 
p(x_{j+1} \mid x_j) = (1-p) \mathcal{N}(x_j, \epsilon) + p \mathcal{N}(x_j, \sigma^2). 
\end{equation}
Here $p\approx 0.01$ can be viewed as the small probability of a CNV event (jump in copy number), while $\sigma^2\gg \epsilon$ ensures that most transitions are small but CNVs lead to significant jumps. Although a piecewise-constant model (where $\epsilon=0$) might be more biologically justifiable, the backwards algorithm described later requires a non-atomic distribution. Instead, we allow for small fluctuations in copy number estimates while preserving the ability to detect abrupt shifts. A visualization of random trajectories sampled from this processs is given in the top left of Figure \ref{figmethod}.

We model the noise in the observed lCNRs using a Laplace likelihood
\begin{equation}
    p(y_j \mid x_j) = \frac{1}{2b}\exp\Bigl(-\frac{\lvert y_j-x_j\rvert}{b}\Bigr).
\end{equation}
The Laplace distribution is heavier-tailed than the more common Gaussian distribution, preventing outliers due sequencing artifacts and other sources of noise from overly influencing inference of the true lCNRs. Finally, we impose a weak prior on the first $lCNR$ of $p(x_0)=\mathcal{N}(0,\tau)$, corresponding to a weak assumption that it is copy-neutral. The complete state-space model is
\begin{equation}\label{eq:state_space_model} 
\begin{split} 
p(x_0) &= \mathcal{N}(0, \tau), \\ 
p(x_{j+1} \mid x_j) &= (1-p) \mathcal{N}(x_j, \epsilon) + p \mathcal{N}(x_j, \sigma^2), \\ 
p(y_j \mid x_j) &= \text{Laplace}(x_j, b). 
\end{split} 
\end{equation}

\subsection{Inference Via Sequential Monte Carlo}
\label{ssec:sequential_mcmc}

The state-space model in \eqref{eq:state_space_model} does not provide a simple closed-form expression for the posterior distribution of $p(x\mid y)$, the lCNR given the observed counts. A common solution in Bayesian inference is to use Monte Carlo methods \cite{robert2004monte}, which generate samples from the posterior distribution to approximate key quantities such as means, medians, and quantiles. When direct sampling is intractable, a Markov chain can be constructed that converges to the posterior distribution. However, standard Markov Chain Monte Carlo (MCMC) methods \cite{betancourt2017conceptual} perform poorly for state-space models due to high correlations in the latent variables, making inference unreliable for all but the shortest sequences \cite{chopin2020introduction}. Sequential Monte Carlo, or particle filtering, overcomes these limitations by using the Markov property of state-space models,where $x_{j+1}$ depends only on $x_j$, to improve inference quality significantly. There are three components we discuss: the forward filtering algorithm, the backward sampling algorithm, and auxiliary filtering.

\subsubsection{Forward Filtering Algorithm}

The forward-filtering step approximates $p(x_j \mid y_{0:j})$ using a custom form of importance sampling \cite{goodfellow2016deep}. The algorithm is
\begin{enumerate}
    \item Initialization: Generate $N$ particles from the prior distribution: $X_0^N \sim p(x_0)$.
    \item Weighting: Assign importance weights based on how well the data align with each particle $w_0^N=p(y_0|X_0^N)$
    \item Auto-normalize the weights: Make the average value of the weights sum to 1 as $W_0^N=w_0^N/\sum_{i=1}^N w_0^i$
    \item Resampling: Select particles according to their weights, favoring those with higher likelihoods
    \begin{equation}
        A_0^{1:N}=\text{resample}(W_0^{1:N})
    \end{equation}
    \item Propogation: Generate new samples using the transition kernel:
    \begin{equation}
        X_1^{1:N}\sim p(x_1|X_0^{A_0^N})
    \end{equation}
    \item Iteration: Repeat this process for all time steps $J$, producing a sequence of weighted samples $(X_j^N, W_j^N)$ that approximate each posterior distribution.
\end{enumerate}
Like importance sampling, any quantity can be computed as a weighted sum of the function applied to the individual samples. However, unlike  standard importance sampling it incorporates a resampling step. Traditional importance sampling estimates quickly become effectively based on a single ``best'' sample. While resampling increases the variance of the estimate in the short term, by ``pruning'' the bad samples it allows for multiple good samples to be propagated to the next element in the sequence.  

\subsubsection{Backward Smoothing for Posterior Estimation}

The forward algorithm allows us to characterize $p(x_j \mid y_{0:j})$, which means the estimates only use information in the spatially preceding amplicons. Ideally, we would like the estimates to use the information of all amplicons in our estimate of the lCNR, particularly for the initial amplicons.This distribution, $p(x_j \mid y_{0:J})$, is significantly more challenging and requires an additional step called backward smoothing to approximate. Intuitively, backwards sampling starts at the end of the sequence and tracks where the origins of the final samples recursively. The algorithm for a single trajectory is 
\begin{enumerate}
    \item Initialize at the final time step $J$. The posterior $p(x_J|y_{0:J})$ is approximated by the forward filter weights $W_J^N$. 
    \item Sample the final point as $B_J\sim \text{Mult}(W_J^{1:N})$.
    \item Compute the backward weights $\hat{w}_{J-1}^n=W_{J-1}^np_J(X_J^{B_J}|X_{J-1}^n)$
    \item Auto-normalize the weights $\hat{W}_{J-1}^n=\hat{w}_{J-1}^n/\sum_{m=1}^N\hat{w}_{J-1}^m$ for all $n$
    \item Sample $B_{J-1}\sim \text{Mult}(\hat{W}_{J-1}^n)$
    \item Iterate until $T=0$. 
\end{enumerate}
From this we obtain a single simulated trajectory $(X_0^{B_0},\dots,X_J^{B_J})$. To approximate the posterior we generate $M$ trajectories, giving a total computational complexity of $\mathcal{O}(JNM)$. We use the common choice $M=N$, and once these $N$ trajectories are obtained we can obtain an estimate of the posterior mean as 
\begin{equation}
    lCNR_j = \frac{1}{M}\sum_{m=1}^M X_j^{B_j^m}.
\end{equation}

\subsubsection{Guided Particle Filtering}

The propagation step in the forward filter using the transition kernel is the most straightforward approach (known as the bootstrap filter). However, it performs poorly with this particular transition kernel. Given that we want to the parameter $p$ to be small (CNVs are infrequent), most of the samples for $x_{j+1}$ will be close to the current value $x_j$. When a transition occurs at the edge of the CNV, most of these samples will be far away from the new value, and as a result inference will effectively be based on a small number of samples.

To address this issue, we use guided particle filtering (also known as auxiliary particle filtering, APF). This approach introduces a proposal distribution that selectively generates particles in regions more likely to match the observations. Instead of directly sampling from the transition model, we introduce an alternative proposal density, denoted as $q(x_j \mid x_{j-1})$, with more desirable properities. As long as this altered proposal distribution is ``inverted'' by changing how the weights are computed, inference remains unaltered.

In this work, we chose a transition density that generates the majority of the particles near the previous lCNR estimate $x_{j-1}$. However, we also generate a small proportion of samples near the current empirical lCNR $y_j$. This means that when an abrupt change does occur, there will be some samples near the new value. The guide we used that incorporates these assumptions is a Gaussian mixture model 
\begin{equation}\label{eq:guide} 
q(x_j\mid x_{j-1},y_j)=0.8\mathcal{N}(x_{j-1},\sigma^2) + 0.2\mathcal{N}(y_j,\sigma^2).
\end{equation}
The weights were chosen heuristically such that most samples were likely to be at the current value but sufficient samples would properly explore regions associated with jumps.


\subsubsection{Thalassemia Profiling via Bayesian Decision Theory}  
\label{ssec:profiling}

So far, our focus has been on the first task of estimating the lCNR at each amplicon. However, our ultimate goal is to assign both an alpha thalassemia and beta thalassemia profile to the sample. To accomplish this, we develop a probabilistic method for determining the profile, denoted $D$, by viewing the predefined profiles as the expected value of the lCNR, that is $d_{ij} = E[x_j \mid D = i]$. We assume that the observational likelihood is Gaussian, that is $p(x_j \mid D = i) = \mathcal{N}(d_{ij}, \sigma^2)$, where $\sigma^2$ is a tuning parameter controlling how informative a single amplicon is. Large values of $\sigma^2$ correspond to low confidence in any particular amplicon for determining the profile, small values instead view each amplicon as highly informative. In this portion of the classifier, we will ignore the spatial dependence of the amplicons. Given that $\sigma^2$ is a tuning parameter controlling the information provided by a single amplicon, full modeling of the spatial correlations is redundant.

Under these assumptions, if the true copy numbers were known, the optimal profile assignment could be determined using Bayesian decision theory as
\begin{equation}\label{eq:optimal}
\hat{D} = \arg\max_{i} \prod_{j=1}^J p(x_j \mid D = i) p(D = i).
\end{equation}  
However, we do not observe $x_j$ directly; we only obtain posterior samples given the empirical log ratios. However, we can integrate out the true copy ratios as
\begin{equation}
p(D = i \mid y_{1:J}) = \int p(D = i \mid x_{1:J}) p(x_{1:J} \mid y_{1:J}) dx_{1:J}.
\end{equation}  
This distribution can be approximated using the samples provided by the backward smoothing step. The final profile assignment is then simply selecting the most probable profile. A visualization of the output of our aggregate caller is shown in the bottom row of Figure \ref{figmethod}.

\subsection{Sample Quality Evaluation via Model Evidence}
\label{ssec:quality_score}

Finally, a key concern in many clinical applications is determining whether a sample meets the quality standards necessary for making a reliable diagnosis. It is important to note that this is distinct from having an ``uncertain'' high-variance posterior. Intuitively, an ambiguous opinion after observing high-quality data is very different from ambiguity stemming from low quality data. Any probabilistic model, such as ours defined in \eqref{eq:state_space_model}, provides a natural way to assess the quality of the data as measured by how ``realistic'' the observed data are according to the generative model via the normalization constant of the model. This constant, also known as the Bayesian evidence, can be used for this purpose in any Bayesian model. In spite of its generality, it is rarely used as reliably estimating the Bayesian evidence is difficult. Fortunately, the particle filter offers a straightforward estimator: the mean of the unnormalized importance weights, given by $\ell_j = \frac{1}{N}\sum_{i=1}^N w_i$.

The total normalization constant for a sample is then computed as the product of these individual normalization constants $L=\prod_{j=0}^J \ell_j$. While $L$ provides a measure of model fit, its absolute value is arbitrary and depends on the units of the parameters. Additionally, simple parameter choices, such as $b$ or $\sigma^2$, can significantly influence these values. However, $L$ is useful for comparing samples relative to other samples. Low values compared to other samples indicate that a sample is less compatiable with the model, while high values indicate a better model fit. This provides a natural quality control mechanism—samples with low likelihood scores should be flagged and returned to the clinician for manual review.

As previously noted, high-quality samples containing homozygous deletions can violate model assumptions (namely homozygous deletions leading to 0 counts), leading to artificially low values of $L$. To mitigate this issue, we use a more robust metric: the median of the individual likelihood values
\begin{equation} 
QC_i = \text{Median}({\ell_j}_{j=0:J}). 
\end{equation}
Based on the properties of the median, this will ignore the subset of values violating model assumptions \cite{ronchetti2009robust}, allowing us to evaluate sample quality based on a typical amplicon. We set a threshold for rejecting samples: those with a $QC_i$ score more than three standard deviations below the mean are considered low quality and excluded from further analysis.

\section{Synthetic Results: Evaluating Guide Effects}\label{sec:synthetic}

In this section, we focus on evaluating the impact of our guided filter as compared to the standard bootstrap filter. Traditionally, the effective sample size (ESS), which estimates the number of ``independent'' samples generated by the MCMC algorithm, is commonly used to evaluate performance. However, we introduce additional metrics, namely the root mean squared error (RMSE) of the CNV estimates and the estimated Bayesian evidence.  These provide a more comprehensive comparison between the two inference algorithms, as they measure the quantities produced by our model.

To generate synthetic data, we simulated a panel with 160 amplicons containing two CNVs: one homozygous duplication and one heterozygous deletion, along with 40 control amplicons. We then fit two models: one using the standard bootstrap filter and the other employing the guided particle filter. Both models used 500 particles with identical state-space model parameters and the metrics were evaluated on each. Each experiment was repeated 10 times to compute confidence intervals for all metrics.

\begin{figure}[ht]
\centering
\includegraphics[width=1.0\textwidth]{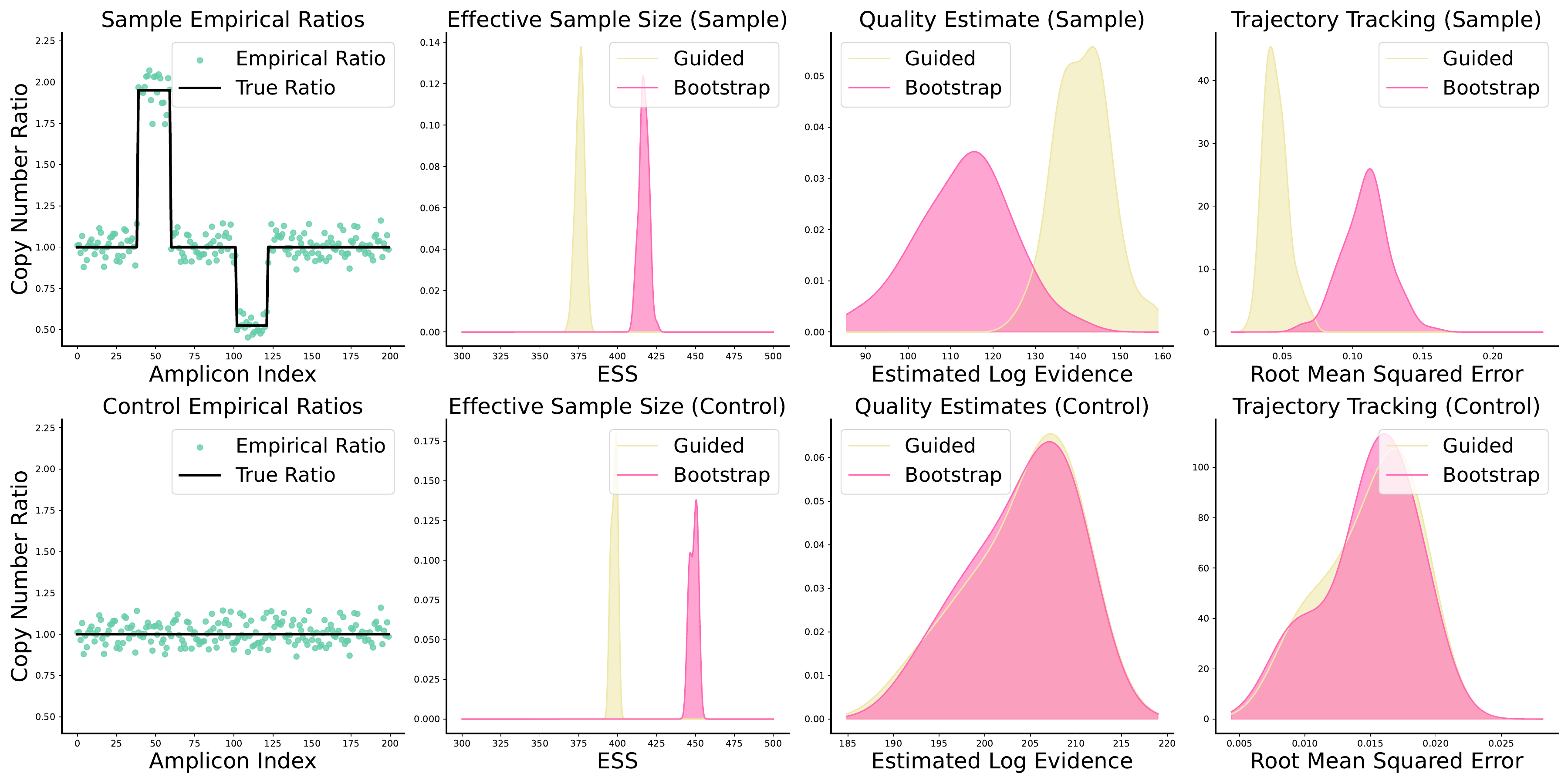}
\caption{The top left shows the empirical CNR from a single synthetic sample with two CNVs, along with the true CNR. The plot in the middle left shows the distribution effective sample size of guided filter in yellow and bootstrap filter in red. Middle right shows the distribution of the log evidence of the sample. Far right shows the RMSE of the posterior median predicting the CNR. The bottom row repeats this process but with a synthetic sample lacking CNVs.}\label{fig1}
\end{figure}

The results for a representative trial are visualized in Figure \ref{fig1}. As expected, the guided filter produces a lower effective sample size of $375\pm5$ compared to $417\pm5$ for the bootstrap filter. Despite this, it substantially improves model fit, particularly in detecting transition points. This is reflected in the likelihood scores: $141.4\pm12.6$ for the guided filter versus $113.4\pm21.8$ for the bootstrap filter. The improved fit is further supported by a significantly lower RMSE ($0.045\pm0.017$) for the guided filter compared to $0.109\pm0.028$ for the bootstrap filter.

To ensure that the guided filter's ability to track abrupt changes in copy number did not degrade performance on normal samples lacking CNVs, we also analyzed a synthetic  ``normal''  (bottom row of Figure \ref{fig1}). The guided filter again showed lower sample size with $398\pm3$ as opposed to $449\pm 4$ for the boostrap filter. However, the remaining metrics were largely unaltered; evidence estimates for the guided filter were ($204.1\pm10.8$) as opposed to $204.1\pm10.4$ for the bootstrap, and the rMSE values were similarly close $0.0155\pm0.006$ versus $0.0149\pm0.006$. While we expect the bootstrap filter to yield marginally higher evidence due to its increased number of samples centered at the current value, the guided sampler does not introduce any statistically significant degradation in performance. Beyond the validation of our inference algorithm, these results illustrate that ESS is not always the most pertinent metric of performance in MCMC.

\section{Cohort}\label{sec:cohort}

\subsection{Data Processing}

We first performed read-to-genome alignment to the GRCh37/hg19 genome using BWA-MEM \citep{li2013aligning}. The initial alignment is improved by performing local realignment using Smith-Waterman \citep{smith1981identification} and a proprietary algorithm. To maximize the base accuracy and minimize the sequencing noise, paired-end reads are assembled into consensus reads, weighted with the base quality scores from both mates. The assembled reads correspond to the gene-specific positions on the genome. A set of filters are applied to remove non-uniquely mapped reads (e.g., pseudogenes) and reads that do not match the amplicon positions (e.g., primer-dimers or non-specific amplifications). For CNV calculations, per-amplicon read depth is calculated for each sample. Counts of overlapping amplicons are stored as independent entries.

\subsection{Thalassemia Samples}

We obtained 44 positive and 13 negative clinical samples as de-identified remnants from a laboratory that tested the samples for clinical use and would have otherwise discarded them. FDA guidance states that such samples may be used for research purposes without obtaining patient consent. These samples were sequenced using a panel of 131 amplicons that covers alpha, beta, delta, gamma and epsilon Thalassemia regions, as well as pseudogenes and control regions. Sequencing was performed on an Illumina MiSeq$^{\mathrm{TM}}$ platform to an average depth of approximately 2800 paired-end reads. The set of true alpha thalassemia diagnoses were obtained from an orthogonal, clinically validated test that uses MLPA with reflex to GAP-PCR for alpha-thalassemia deletion and duplication detection. Of the positive samples, 31 had a heterozygous alpha-3.7 deletion, 5 had a homozygous alpha-3.7 deletion, 2 had alpha-4.2 deletions, 3 had south-east Asian deletions, 1 sample with an alpha-4.2 duplication, 1 sample with a large HBA duplication, and 1 sample with an alpha-3.7 deletion/4.2 duplication.


\subsection{BRCA Samples}

To further investigate our quality control metrics, we obtained a second dataset sequenced using a panel designed to detect mutations in the BRCA1 and BRCA2 genes. This panel contains 283 amplicons targeting various regions within BRCA1 and BRCA2 genes, as well as 8 control amplicons from other chromosomes. We obtained four mutation-positive DNA cell line samples from the Coriell Institute for Medical Research with known mutations: NA18949 with a BRCA1 exon 14 and 15 deletion, NA14626 with a BRCA1 exon 12 duplication, NA0330 with a whole-gene duplication of BRCA2, and NA02718 with a whole gene deletion of BRCA2. We also obtained four negative cell line samples NA12878, NA19240, NA24385, and NA24143. Additionally, Horizon Discovery's Mimix$^{\mathrm{TM}}$ Quantitative Multiplex, fcDNA (Moderate) Reference Standard was used to evalue performance on Formalin-compromised DNA (FFPE). Libraries were prepared and sequenced on Illumina's MiSeq$^{\mathrm{TM}}$. All samples were collected under informed consent. 

\section{Results} \label{sec:results}

\subsection{Overall Study Objectives} 
There are two overall objectives in this section; first, we wish to evaluate the accuracy of our thalassemia profiling and second, we wish to evaluate the ability of Bayesian evidence to recognize substandard samples. To evaluate profiling accuracy, we use the thalassemia dataset from Section \ref{sec:cohort}. To our knowledge, we have no competitor methods to evaluate the performance of our entire informatics pipeline. Instead, we compare the state space portion of the method against other standard smoothing methods. To evaluate the efficacy of Bayesian evidence for quality control, we use data from the BRCA panel. The  process of fixing DNA with formalin when generating FFPE samples causes DNA degradation, leading to erratic empirical copy number ratios when an unmatched (non-FFPE) cell line normal is used as a reference for an FFPE sample. Thus, the FFPE samples in this dataset serve as known ``poor-quality'' samples, allowing us to evaluate the quality score’s effectiveness in distinguishing between cell line and FFPE samples.

\subsection{Evaluating Profile Accuracy and Amplicon-Level CNV Estimation}
\label{ssec:profile_accuracy}

We first evaluate our ability to both detect thalassemia as well as classify thalassemia profiles. To measure this, we evaluated both the sensitivity and specificity of a positive test as opposed to negative, and the accuracy of the profile choice. We were able to increase sample size by seperately considering diagnoses of the alpha and beta status, and by repeating the experiment with three different sets of normal samples for comparison. With this procedure, we obtained 324 measurements, of which 132 were positive and 192 were negative. We were unable to find any other publicly available profilers to provide a full bioinformatic comparison of our method. However, were able to compare our state space model to other methods for denoising the copy number estimates, providing a comparison of the first component of our profiler. We chose comparisons of no smoothing (empirical), kernel smoothing \citep{wasserman2006all}, smoothing splines \citep{hastie2009elements}, and anisotropic diffusion \citep{black1998robust}. We compare our method both when all samples are included (no QC) as well as the performance when subpar samples are removed (QC) according to the procedure described in Section \ref{ssec:quality_score}. The results are shown in Table \ref{tab:model_comparison}. 

We found that even without quality control, our method had the highest accuracy of all considered methods with an accuracy of $91.5\%$ as compared to the next highest (empirical with $91.2\%$). This further rises to $93.9\%$ when substandard samples are excluded. Our method has lower specificity than smoothing splines $0.93$ as opposed to $0.94$, but this is rectified by quality control, which raises the specificity to $0.96$, the best among all considered methods.

\begin{table}[ht]
\centering
\begin{tabular}{l|cccccccc}
\textbf{Model} & \textbf{Accuracy} & \textbf{TP} & \textbf{FP} & \textbf{FN} & \textbf{TN} & \textbf{Sens.} & \textbf{Spec.} & \textbf{RMSE}\\
\midrule
Empirical & 91.2\% & 131 & 20 & 1 & 172 & 0.99 & 0.90 & 0.156 \\
Kernel Smoothing & 90.9\% & 131 & 19 & 1 & 173 & 0.99 & 0.90 & 0.133 \\
Smoothing Splines & 90.4\% & 129 & 12 & 3 & 180 & 0.98 & 0.94 & 0.135\\
Anistropic Diffusion & 90.1\% & 131 & 19 & 1 & 173 & 0.99 & 0.90 & 0.131 \\
Ours (No QC) & 91.5\% & 131 & 14 & 1 & 181 & 0.99 & 0.93 & 0.131\\
\textbf{Ours (QC)} & \textbf{93.9}\% & 128 & 8 & 1 & 178 & 0.99 & \textbf{0.96} 
 & \textbf{0.111} \\
\midrule
\end{tabular}
\caption{Comparison of predictive performance of the state space model, both with quality control on the quality score (QC) and without such filtering (No QC), as compared to other smoothing methods}
\label{tab:model_comparison}
\end{table}

We also evaluate the ability of our model to estimate the CNR on an amplicon-by-amplicon basis, as specific amplicons or exons may be of interest in addition to predefined profiles. We can reuse the experimental setup from the previous section, except now we will compare the output of the state space model to evaluate how close the CNR is to the true ratio. We found that without QC our method matched the root mean squared error (rMSE) of the best smoothing method (anisotropic diffusion). However, once low-quality samples were excluded, the RMSE of our method was superior with a value of $0.111$.

\begin{figure}[ht]
\centering
\includegraphics[width=1.0\textwidth]{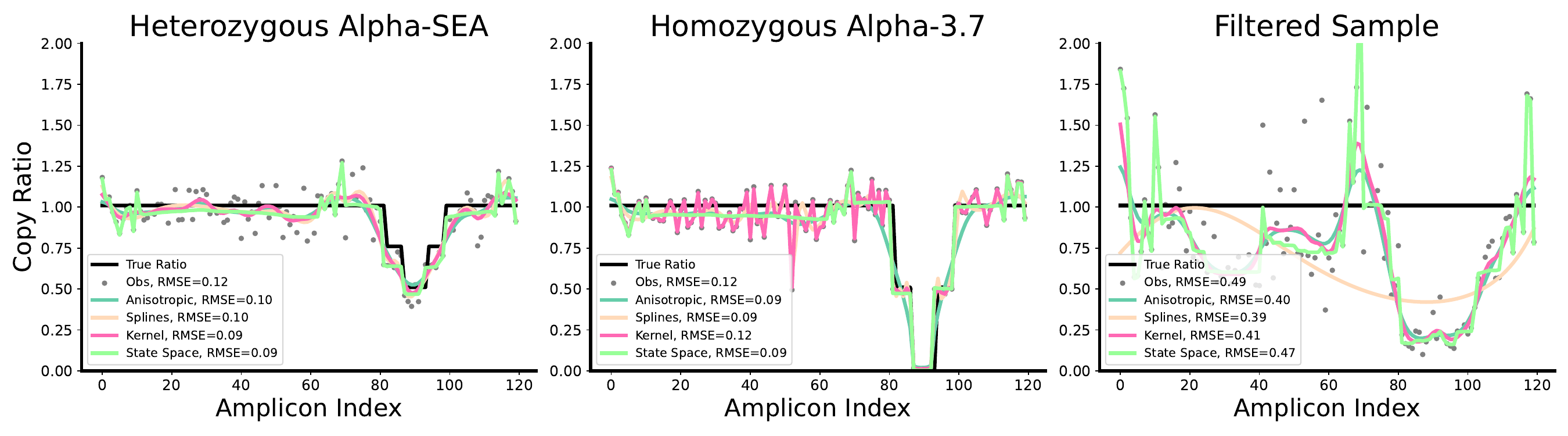}
\caption{The results of the different smoothing methods on three different samples. The third plot shows a sample that was rejected by our quality control criterion.}\label{fig:jointplot}
\end{figure}




\subsection{Evaluating the Sample Quality Metric}
\label{ssec:quality}

\begin{figure}[h]
\centering
\includegraphics[width=1.0\textwidth]{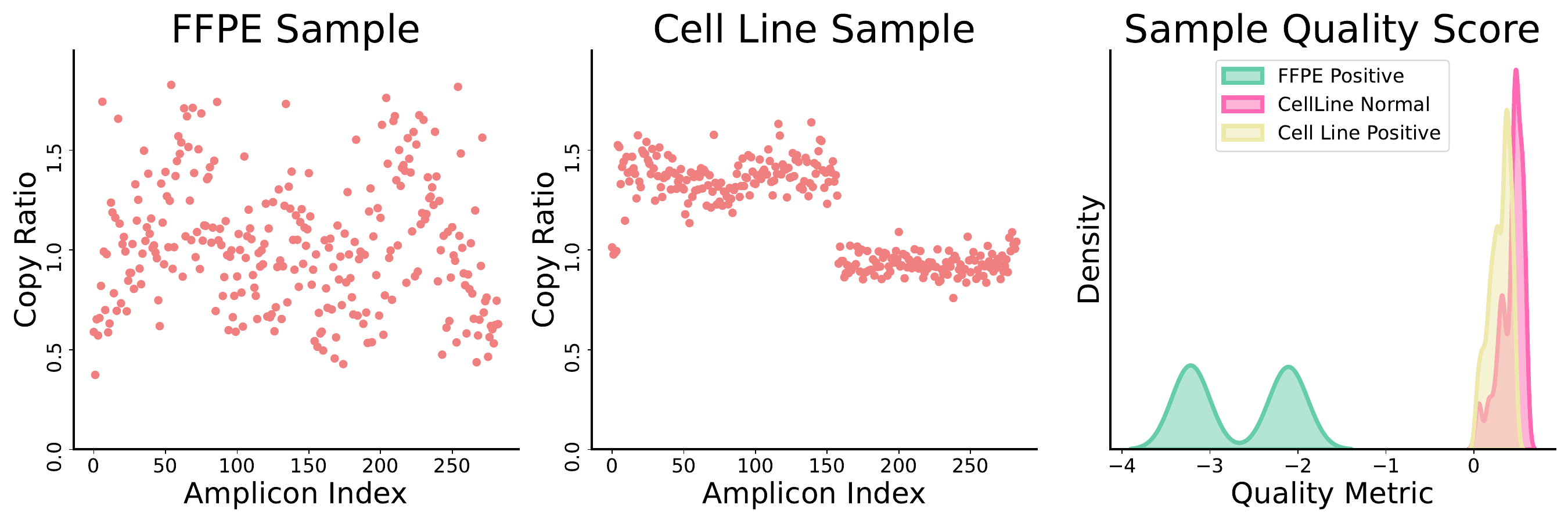}
\caption{This figure visualizes the sample quality differences detected by our metric. On the left, we show the CNRs of an FFPE sample with a cell-line normal. These ratios fluctuate wildly between $0.5$ and $1.5$, with a standard deviation of $0.6$. In comparison, the ratio of a cell line positives with the same normal shown in the middle plot has substantially reduced noise in the CNRs, with a standard deviation of $0.16$. Finally, on the right we visualize the QC score standardized by batch to remove run-specific variability. We can see that the FFPE positive samples are all below $-2.0$, while both positive and normal cell-line samples were above $0$.}\label{fig:samplequality}
\end{figure}

Finally, we evaluate the ability of our model to detect poor-quality samples for quality control. While we already have evidence for its efficacy due to it improving specificity in Section \ref{ssec:profile_accuracy}, here we quantify its performance in differentiating low-quality samples from higher quality samples. We fit our state space model on each sample compared to three of the Coriell normals, evaluated the log evidence, and then calculated each sample's quality score. The results are shown in Figure \ref{fig:samplequality}. On the left, we show the ratios associated with one of the FFPE samples. We can see that the empirical copy ratios are far more erratic in the FFPE samples as compared to the clinical sample shown in the center. This is reflected in the quality scores obtained; the average score for the TV samples and clinical samples  are $0.42\pm 0.08$ and $0.29\pm0.08$ respectively while the scores for the FFPE samples are $-2.66\pm0.8$. From this, we can see a clear separation between the a priori ``low-quality'' samples and the samples that match model assumptions. Based on the quality metric introduced in the methods, all samples from FFPE would have failed quality control and failed to produce a call. By contrast, only a single sample from the cell line positive would have been falsely rejected as being substandard.

\section{Discussion} \label{sec:discussion}


From a clinical perspective, this study makes two key contributions. First, to our knowledge, this is the first open-source thalassemia profiler using targeted sequencing. Furthermore,  given the widespread prevalence of CNV disorders, this tool has significant clinical applications for other disorders beyond thalassemia, including the BRCA variations mentioned in \ref{sec:cohort}. Second, our work demonstrates the utility of Bayesian evidence in identifying low-quality samples, enabling clinicians or technicians to flag them for manual review or rerunning. 

From a machine learning perspective, our work highlights the importance of auxiliary particle filtering in detecting and characterizing change points using state-space models. In regions with dramatic shifts, the standard bootstrap filter tends to under-sample critical areas, leading to reduced predictive accuracy and an underestimate of the Bayesian evidence in the sample. Traditional metrics, such as the average effective sample size, fail to capture this issue and may misleadingly suggest the superiority of the bootstrap filter.

\paragraph{Limitations} While our method performs well, there are several remaining limitations. First, our evaluation was limited both in sample size and variety in profiles. A larger cohort, including the more obscure variants, would be highly beneficial in improving parameter choices and performance evaluation. Second, our method currently requires incorporation of an in-batch normal, which can increase costs of diagnosis. Future work will explore alternative strategies, such as direct modeling of amplicon efficiency, allowing for true normal-free calling, or alternatively incorporation of a ``panel of normals'' strategy designed to repeatedly reuse large numbers of normals in new runs. 


\bibliography{main}


\end{document}